\documentclass[12pt]{article}
\usepackage{amsmath,amssymb}
\newcommand{\beq}{\begin{equation}}
\newcommand{\eeq}{\end{equation}}
\newcommand{\ba}{\begin{array}}
\newcommand{\ea}{\end{array}}
\newcommand{\bea}{\begin{eqnarray}}
\newcommand{\eea}{\end{eqnarray}}
\newcommand{\bean}{\begin{eqnarray*}}
\newcommand{\eean}{\end{eqnarray*}}

\newenvironment{pf}{\begin{proof} \rm}{\hfill$\square$ \end{proof}}
\newcommand{\CM}{{ M}}

\makeatletter
\@addtoreset{equation}{section}

\makeatother
\def\la{\lambda}

\newcommand{\rref}[1]{(\ref{#1})} %puts parentheses around ref's

\def\mat2#1#2#3#4{{\left(\begin{array}{cc}#1 & #2\\ #3 & #4
      \end{array}\right)}}

\def\mats2#1#2#3#4{{\left(\begin{array}{cc}#1 & #2\vspace{2truemm} \\ #3 & #4 
\end{array}\right)}}
\def\vec2#1#2{{\left[\begin{array}{c}
#1 \\ #2 \end{array}\right]}}

\def\ddd#1#2{\displaystyle{\frac{\partial #1}{\partial #2}}}

\def\bih{bi-Ham\-il\-tonian}

\begin{document}
\begin{center}{\Large\bf 
Bi--hamiltonian Geometry and Separation of\\\vspace{.3truecm} Variables
for Gaudin Models: a case study}\\
\vspace{1.0truecm}
Gregorio Falqui and Fabio Musso
\\
SISSA, Via Beirut 2/4, I-34014 Trieste, Italy\\
falqui@sissa.it, fmusso@sissa.it
\end{center}
\vspace{0.5truecm}
\abstract{ We address the study of the classical Gaudin spin model 
from the \bih\ point of view. We describe in  details the $sl(2)$ 
three particle case}

\section{Introduction}
The Gaudin models are among the best known integrable lattice
models~\cite{Gaudin1,Gaudin2}. Physically, they consist of a system of
$N\, sl(2)$--valued ``spins'' $A_i$, possibly coupled to an external 
magnetic field  whose strength is assumed to depend on the lattice site. 
Its phase space $M$ can be identified with  the
n--fold product of $sl(2)$, and the  Hamiltonian is
\begin{equation}
  \label{eq:i1}
  H_G=\frac12 \sum_{i\neq j =1}^N {\rm Tr}(A_iA_j)+\sum_{i=1}^N  {\rm Tr}
(a_i \sigma A_i), \qquad \sigma, A_i\in sl(2)\>.   
\end{equation}
On $M$ there is a natural Hamiltonian structure, the 
direct product of the Poisson--Lie bracket on $sl(2)$. 
Complete integrability can be deduced by considering the Lax representation of
such a model. One considers the Lax matrix
$  L_{rat}(\la)=\sigma+\sum_{i=1}^N \frac{A_i}{\la-a_i}$,
and proves\cite{RSTS,Ha1,Ha2} that \\
a) The Hamiltonian flow associated with $H_G$ is a Lax flow;\\
b) integrals of the motion are provided by the spectral 
invariants of $L$;\\
c) the Lax equations admits an $R$--matrix formulation, so that the integrals
of the motion are in mutual involution.

In this note we want to address the problem from the standpoint of \bih\
geometry, namely to frame 
these models  within the so--called 
Gel'fand--Zakharevich (GZ)~\cite{GZ} set--up for \bih\ integrable systems. 
Essentially, the core of such an approach relies in the extensive use of
the Lenard recursion relation, associated with a \bih\ structure, 
to generate mutually commuting constants of the
motion, via the so--called method of the Casimirs of the Poisson 
pencil. 

To equip the Gaudin model with a \bih\ structure we will use a suitable
strategy, to be discussed in Section~\ref{sec:model}, which consists in
considering, along with the Lax matrix $L_{rat}$ a {\em polynomial} Lax matrix
$L_{pol}=\prod_{i=1}^N (\la-a_i) L_{rat}$, and recalling that
on the space of  polynomial matrices a family of
mutually compatible Poisson structures is defined. 
The natural Poisson structure we referred above
turns out to be a specific element of such a family, and so 
we can define a second Poisson 
structure for the Gaudin model simply choosing 
another element of such a family. Successively, we use this \bih\ structure 
to recover complete integrability of the model following the GZ recursion
procedure. Finally, we will briefly address the problem of
Separation of variables, in the \bih\ set--up discussed in\cite{fp1}.
 
Such an analysis can be performed for an arbitrary number of spins (or
``particles''), and, {\em mutatis mutandis} letting the ``spins'' be $sl(n),
n\ge 3$ valued. A full account of this program is
outside the size of this paper. 
In this note we will use the three particle $sl(2)$ model for a case
study.  The above mentioned generalization of this approach 
will be published elsewhere. 

\section{The model and the \bih\ structure}\label{sec:model}
We consider the three-particle $sl(2)$ Gaudin model, 
whose degrees of freedom are encoded in the Lax matrix with spectral parameter
$\la$
\begin{equation}\label{eq:i2}
L_{rat}=\sigma+\sum_{i=1}^3 \frac{A_i}{\la-a_i},
\end{equation}
where the $a_i$ are distinct constant parameters and 
\begin{displaymath}
A_i= \left(
\begin{array}{cc}
 h_i/2 & f_i \\ 
e_i & -h_i/2
\end{array}
\right),
\> i=1,2,3,\quad \sigma=\left(
\begin{array}{cc} 1&0\\
    0&-1\end{array}\right).
\end{displaymath}
The phase space $\CM$ is\footnote{Hereinafter we will not distinguish
  between $sl(2)$ and its dual, assuming implicitly  to identify the two 
  by means of the Killing form.} the Cartesian product of three copies of
$sl(2)$;
on $\CM$ one can define ``standard'' Poisson brackets, simply taking the
Cartesian product of the Lie--Poisson brackets on $sl(2)$:
\begin{displaymath}
\{ h_i, e_j \}=2 \delta_{ij} e_i \qquad \{ h_i, f_j \}=-2 \delta_{ij} f_i
\qquad \{ e_i, f_j \}= \delta_{ij} h_i
\end{displaymath}
Here and in the sequel we will extensively use  a ``matrix'' representation of
these brackets (rather, of the Poisson tensor $P$ 
associated with these brackets),
which can be explained as follows. 

We identify the tangent and cotangent bundles to $\CM$ with $\CM$ itself, so
that a tangent vector will be written as the vector $\dot
X=\left(\dot{A_1},\dot{A_2},\dot{A_3}\right)$ 
and the differential of a function $F$ as
$dF=\left(\ddd{F}{A_1},\ddd{F}{A_2},\ddd{F}{A_3}\right)$, 
and notice that the Hamiltonian
vector fields associated with the Lie--Poisson brackets are given by
$
\dot{A_i}=[A_i,\ddd{F}{A_i}],\> i=1..3.$
We can recast in matrix form such a formula as: 
\begin{equation}\label{eq:poidiag}
\dot{X}=\left( 
\begin{array}{c}
\dot{A_1}\\
\dot{A_2}\\
\dot{A_3}  
\end{array}
\right)=
P dF= 
\left( 
\begin{array}{ccc}
[A_1,.] & 0 & 0\\
0 & [A_2,.] & 0\\
0 & 0 & [A_3,.]  
\end{array}
\right)\cdot 
\left( 
\begin{array}{c}
\ddd{F}{A_1}\\
\ddd{F}{A_2}\\
\ddd{F}{A_3}
\end{array}
\right) 
%\label{PDA}
\end{equation}
Similarly, we will write in an analogous matrix form the linear Poisson
brackets we are going to discuss in the sequel.

To endow $\CM$ with a \bih\ structure
consider, along with the rational Lax matrix $L_{rat}$ the
{\em polynomial} Lax matrix 
\begin{equation}
  \label{eq:1.3}
  L_{poly}=\prod_{i=1}^3 (\la-a_i)L_{rat}=\la^3\sigma+B_2\la^2+B_1\la+B_0.
\end{equation}
Explicitly, this change of coordinates on $M$ is given by: 
\begin{eqnarray}\label{eq:trs}
\nonumber && B_0=a_2 a_3 A_1+a_1 a_3 A_2 + a_1 a_2 A_3 -s_3 \sigma\\
&& B_1=-(a_2+a_3) A_1-(a_1+a_3) A_2 - (a_1+a_2)A_3 +s_2
\sigma\\
\nonumber && B_2=A_1+A_2+A_3-s_1 \sigma
\end{eqnarray} 
with
\begin{equation}\label{eq:sympol}
 s_1=(a_1+a_2+a_3), s_2=a_1 a_2 +a_1 a_3+ a_2 a_3,\> 
s_3= a_1 a_2 a_3\> .
\end{equation}

The rationale for this ``coordinate change'' is that,
on the space of polynomial pencils of matrices a family of
mutually compatible Poisson brackets are defined\cite{RSTS} via R--matrix
theory. In a nutshell, this family can be described (in our three particle 
case) by saying that there is a map $\phi$ 
from the space of degree three polynomials 
in the variable $\la$
to the set of Poisson structures on the manifold of polynomial Lax matrices
of the form~\rref{eq:1.3} which sends the monomials $\la^0, \ldots, \la^3$
into four fundamental Poisson brackets, $\Pi_i, i=0, \ldots, 3$.  
These fundamental brackets are written as follows:
\begin{eqnarray*}
%\label{eq:fupoi}
&& \Pi_0=
\left(
\begin{array}{ccc}
 {}[B_1,.] & {}[B_2,.] & {}[\sigma,.] \\
{}[B_2,.] & {}[\sigma,.] & 0 \\
{}[\sigma,.] &0 & 0   
\end{array} \right)
\quad\quad\quad
\Pi_1=\left(\begin{array}{ccc}
 -[B_0,.] & 0& 0\\
0& [B_2,.]&[\sigma,.] \\
0&[\sigma,.] &0   
\end{array} \right)\\ 
&& \Pi_2:=\left(
\begin{array}{ccc}
0& -[B_0,.] & 0\\
-[B_0,.]&-[B_1,.]&0 \\
0&0&[\sigma,.]   
\end{array} \right)
\quad 
\Pi_3=-\left(
\begin{array}{ccc}
0& 0& {}[B_0,.] \\
0& {}[B_0,.]& {}[B_1,.]\\
{}[B_0,.]&{}[B_1,.]& {}[B_2,.]   
\end{array} \right)
\end{eqnarray*}  
A straightforward computation shows that under the map connecting the $A_i$'s
with the $B_j$'s, the standard Poisson bracket~\rref{eq:poidiag} is sent into
the combination:
\begin{equation}\label{eq:poisdb}
\Pi_3-s_1\Pi_2+s_2\Pi_1-s_3\Pi_0
\end{equation}
of the fundamental brackets. So we can regard  
$P$ as being associated by $\phi$ with the polynomial
\[
p(\la):=(\la-a_1)(\la-a_2)(\la-a_3)=\la^3-s_1 \la^2+s_2\la-s_3
\]
and we consider the Poisson tensor 
$Q=\Pi_2-s_1\Pi_1+s_2\Pi_0=\phi(q(\la))$, where 
\[
q(\la)=({p(\la)}/{\la})_+=\la^2-s_1\la+s_2\>.
\]
Translating back in the coordinates associated with the matrices $A_i$ the
tensor $Q$, we get that its components are given by expressions of the form:
\begin{equation}\label{eq:qpoi}
Q_{i,j}=\left(\sum_{k=1}^3 c^k_{i,j}[A_k,.]\right)+d_{i,j}[\sigma,.]
\end{equation}
where the $c^k_{i,j}$ and the $d_{i,j}$ 
are somewhat complicated rational functions of the parameters $a_1,a_2,a_3$,
whose explicit expressions are irrelevant here.

Summing up, we have equipped the phase space $M=\left(sl(2)\right)^3$ 
of the Gaudin model with the Poisson pencil
\[
P_\la=Q-\la P.
\]
Quite clearly, the specific choice of the polynomial $q(\la)$ is somewhat
arbitrary, and different choices could be considered. However, in this paper
we will stick to this choice, and,  in the next Section we will study 
the GZ geometry of such a pencil.
\section{The GZ analysis of the Poisson pencil}
The Gel'fand--Zakharevich method is a ``recipe'' to associate with a Poisson
pencil $Q-\la P$ an integrable system (that is, a family of mutually commuting 
vector fields). It is particularly suited for studying Poisson pencils of
non--maximal rank. Its core
can be described as follows:

Let $M$ be a $2n+k$ dimensional \bih\
manifold, and let us suppose that the rank of $P$ and $Q$ equals $2n$.
One starts fixing a basis $\{H^0_i\}_{i=1\ldots, k}$ of the
Casimir functions of $P$, that is functions satisfying  $PdH^0_i=0$. 
Applying the second tensor $Q$ to one of such functions one gets (in general) 
a non trivial vector field $X_i^1=QdH^0_i$, and tries to find another
function $H_i^1$ such that $QdH^0_i=P dH_i^1$. If such an equation can be
solved for $H^1_i$, then one generates a new vector field
applying $Q$ to $dH_i^1$, and so on and so forth. 
Supposing one can iterate this process, one finds, for
each independent Casimir of $P$, a Lenard chain of vector fields $X^a_j$. 
As a consequence of the \bih\ recursion relations, all functions
$H_i^j$ commute (w.r.t. both Poisson brackets), even if they do not pertain to 
the same Lenard chain. Complete integrability (of every vector field of the
chains) is recovered whenever 
$n+k$ elements of the family of functions $H^j_k$ 
are functionally independent, noticing that $k$ elements of such
``fundamental'' Hamiltonians are provided by the $k$ Casimirs $H^0_i$. Indeed,
all vector fields of the Lenard chains are tangent to
the generic symplectic leaf of $P$, and, when restricted to such a leaf, the
remaining $n$ functions of the fundamental Hamiltonians provide the required
$n$ mutually commuting integrals of the motion.

To apply these ideas to the three particle Gaudin model, we first notice 
that ${\rm dim} M=9$ and the rank both of $P$ and $Q$ equals $6$. Actually,
a basis $H^0_i$ of Casimirs is given by the three functions
\[
H^0_i={\rm Tr} A_i^2={\rm res}_{\la=a_i}(\la-a_i){\rm Tr}(L_{rat}^2)
\]
Furthermore, applying $Q$ to such functions we get three (independent) vector
fields $X_i$. Finally, a long but straightforward computation shows that
it is possible to find three additional functions $H^1_i$ such that:
\[
P d H_i^1=X_i,\qquad Q dH_i^1=0
\]  
In other words, we can arrange these six functions in three Lenard chains of
the form
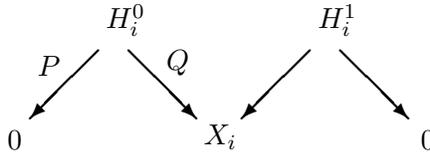
\setlength{\unitlength}{2500sp}
\begin{center}
\thicklines
\begin{picture}(8475,1000)(1426,-500)
\put(4400,250){\vector(-1,-1){675}}
\put(4700,250){\vector( 1,-1){675}}
\put(6500,250){\vector(-1,-1){675}}
\put(6800,250){\vector( 1,-1){675}}
\put(3500,-730){\makebox(0,0)[lb]{$0$}}
\put(7600,-730){\makebox(0,0)[lb]{$0$}}
\put(5450,-730){\makebox(0,0)[lb]{$X_i$}}
\put(4480,390){\makebox(0,0)[lb]{$H^0_i$}}
\put(6580,390){\makebox(0,0)[lb]{$H^1_{i}$}}
\put(3800, 0){\makebox(0,0)[lb]{$P$}}
\put(5075, 0){\makebox(0,0)[lb]{$Q$}}
\end{picture}
\end{center}
The new Hamiltonians $H^1_i$ are a linear combinations of the
Casimirs $H^0_i$ and of the other independent spectral 
invariants $K_i$ of the Lax matrix, given by
\[
K_i={\rm res}_{\la=a_i}{\rm Tr}(L_{rat}^2)=2\left(\sum_{j\neq i}\frac{{\rm Tr}
(A_iA_j)}{a_i-a_j}+ a_i{\rm Tr}(\sigma A_i)\right).
\]
For instance one has
\begin{equation}
{H}_1^1=\frac{(-a_1 g_1 H^0_1+
a_2 a_3(a_2-a_3)(a_2 H^0_2-a_3 H^0_3)+t_1K_1)}{a_1(a_1-a_2)^2(a_1-a_3)^2}
\end{equation}
with 
\[
g_1=a_1(s_1^2-3s_3),t_1=a_1a_2a_3(a_1-a_2)(a_1-a_3).
\]
($H^1_2$ and $H^1_3$ are obtained
by cyclic permutations of the indexes $(1,2,3)$). Notice that there is an
invertible linear relation (with constant coefficients) between the basis of
spectral invariants $\{H^0_i, K_i\}$ and the basis of GZ Hamiltonians $\{H^0_i,H^1_i\}$

Summing up the \bih\ structure we consider
gives rise to three Lenard chains of ``length'' one each (that is each one
comprising one independent vector field). Since the ``physical''
Gaudin Hamiltonian~\rref{eq:i1} is a linear combination of the $K_i$,
$H_g=\frac12\sum_{i=1}^3 a_i K_i$ and hence of the GZ Hamiltonians, we have
recovered complete integrability of the model. Finally, 
notice that we can collect the Hamiltonians $\{H^0_i,H^1_i\}$
into the three polynomials 
\[
F_i(\la)=\la
  H^0_i+{H}_{i}^1,
\] 
and algebraically represent the short Lenard chain(s) 
depicted above by means of the formula
$\left(Q-\la P\right) d F_i(\la)=0$.

\section{On Separation of Variables}\label{sec:sov}
In this last Section we will briefly address the problem of Separation of
Variables for the inhomogeneous Gaudin models.
Separation of variables (SoV)
for this model was proven\cite{Sk89,Sk95,Ha1,Ha2} , 
who defined separated variables as
coordinates of the poles of a suitably normalized Baker--Akhiezer function 
associated with
the Lax matrix~\rref{eq:i2}. A geometrical approach to the SoV problem, based
on \bih\ geometry,  has quite recently been discussed 
in the literature\cite{mt97,Bl98,FMT00,fp1}. Here
we want to show that SoV for the Gaudin models falls within this scheme.

The basic geometrical object underlying  the \bih\ approach to SoV is 
a so--called $\omega N$ manifold, that is, 
the datum of a
symplectic manifold $(V,\omega)$ endowed with an additional $(1,1)$--tensor
$N$ with vanishing torsion satisfying a suitable compatibility condition with
respect to the symplectic 2--form $\omega$ (a {\em Nijenhuis} tensor for
short) The content of the ``\bih'' SoV theorem (see~\cite{fp1})
can be stated as follows:

{\bf I}) On (suitable open sets of) of the $\omega N$ manifold $V$
a family distinguished canonical coordinates (called
Darboux--Nijenhuis (DN) coordinates) is intrinsically defined through 
the spectral properties of its Nijenhuis tensor $N$~\cite{Ma90}.

{\bf II}) A Liouville integrable system, 
characterized by a complete set of mutually
(w.r.t. the Poisson tensor $\omega^{-1}$) integrals of the motion 
$\{H_i\}_{i=1,\ldots, N}$ 
is separable in DN coordinates if and only if these Hamiltonians commute also
w.r.t. the second bracket defined by 
\begin{equation}\label{eq:pomn}
Q=N\cdot \omega^{-1}.
\end{equation}
As we have seen, the Gaudin models (as it is the case for a number of
integrable systems related with R--matrix theory and/or reductions of soliton
equations) admit a (reasonably) natural \bih\ formulation on a \bih\ manifold
where none of the brackets of the pencil is non degenerate. So, the first
problem to be tackled in such an instance is the so--called reduction problem, 
that is to concoct from the (degenerate) \bih\ structure of the problem a
structure of $\omega N$ manifold on suitable submanifolds of $M$, in such a
way to satisfy condition {\bf II} above. 

To show that this can be successfully achieved in the Gaudin models 
we will follow a recipe discussed in~\cite{FMT00,fp1}, specifically suited
for GZ systems.
We consider the Poisson pencil $Q-\la P$, and fix a basis 
$\{H^0_1,\ldots,H^0_k\}$ of 
Casimirs of $P$, where we assume that $k=corank(P)=corank(Q)$.
The symplectic leaves $S_c$ of $P$ are thus characterized 
by the equations $H_i^{0}={\rm const}_i$, 
and come equipped with a natural symplectic structure. It holds\cite{fp1} the
\\
{\bf Proposition:}
In the above geometrical set--up, let $Z_1,\ldots,Z_h$ be a family of vector
fields on $M$, transversal to the symplectic leaves of $P$ such that the
functions vanishing along the distribution $<Z_i>$
generated by the $Z_i$'s are a
Poisson subalgebra both w.r.t. $P$ and $Q$. Then the generic symplectic leaf
of $P$ has the structure of an $\omega N$ manifold. The Nijenhuis tensor $N$
is defined, via Eq.~\rref{eq:pomn} by the restriction of the following
(modified) Poisson tensor $
\widetilde{Q}=Q-\sum_{i=1}^h Q(d(H_i^{top}))\wedge \widetilde{Z_i},$
where $\widetilde{Z_i}$ are a normalized basis (possibly defined on an open
subset of $S_c$) for $<Z_i>$, i.e. $\widetilde{Z_j}( H_i^{0})=\delta_{i j}$.
Condition {\bf II} above is automatically satisfied.\hfill $\square$
\\

We will now apply this scheme to the three particle $sl(2)$ Gaudin model,
endowed with the \bih\ structure $Q-\la P$ introduced in Section 
\ref{sec:model}. As we have seen, 
the GZ structure of the problem
is quite simple. We have three  Casimir polynomials
$F_i(\la)$ of degree one, (meaning that, indeed $corank(P)=corank(Q)=3$) 
and so we have to find a three--dimensional
distribution satisfying the properties of the Proposition recalled
above.

The idea to solve this problem is very simple, and relies on the following
observation on the (ordinary) Lie-Poisson brackets on a single copy of
$sl(2)$. With the notations of
Section \ref{sec:model} the Poisson bracket of two functions $F,G$ on $M$,
is given by
$\{F,G\}={\rm Tr}(\ddd{F}{A}\cdot [A,\ddd{G}{A}])=
-{\rm Tr}(A\cdot[\ddd{F}{A},\ddd{G}{A}]).$
Parametrizing the generic element $A$ of $sl(2)$ as  $
A= \left(
\begin{array}{cc}
h/2 & f \\ 
e & -h/2
\end{array}
\right)
$
we consider the vector field $Z=\ddd{}{e}$. We notice that differentials of
functions vanishing along $Z$ admit a very simple matrix
representation. Indeed $Z$ is matricially represented as $
Z(A)=\left(\begin{array}{cc} 
  0 &0\\ 1 & 0\end{array}\right), $
and so $Z(F)=0$ iff $\left(\ddd{F}{A}\right)_{1,2}=0$, i.e., iff $\ddd{F}{A}$
lies in the lower Borel subalgebra ${\bf b_-}$ of $sl(2)$.

Let $F,G$ be functions such that $Z(F)=Z(G)=0$, and
let us compute $Z(\{F,G\})$. Thanks to the Leibniz property of
the Lie derivative and the fact
that $Z$ is a constant vector field we have that
\begin{equation}\label{eq:zff0}
Z(\{F,G\})=-{\rm Tr}(Z(A)\cdot [\ddd{F}{A},\ddd{G}{A}])
\end{equation}
which vanishes as well since ${\bf b_-}$ is indeed a Lie subalgebra of $sl(2)$.
Finally we notice that the Casimir $C$ of the Lie-Poisson bracket is given by
$C=h^2/2+2ef$, so $Z(C)=2f$ and hence $Z$ is generically transversal to the
symplectic foliation of the Lie Poisson brackets; in particular, the
normalized generator $\widetilde{Z}$ is given by $1/2f Z$.  
   
To use this simple result in the case of the three particle $sl(2)$ Gaudin 
model we consider the vector fields $Z_i=\ddd{}{e_i}, i=1,2,3.$. The
differentials of functions $F$ vanishing along $<Z_i>$ are represented by
triple of matrices $dF=(\ddd{F}{A_1},\ddd{F}{A_2},\ddd{F}{A_3})$ with
$\ddd{F}{A_i}\in {\bf b_-}, \> i=1,2,3$. The fact that such functions are a 
Poisson subalgebra for $P$ is self evident. To ascertain that the same is true 
for $Q$ one simply has notice that, using the explicit
expressions~\rref{eq:qpoi}, 
the brackets $\{F,G\}_Q=<dF,Q dG>$ are given by the multiple sum
\[
\{F,G\}_Q=\sum_{i,j}{\rm Tr}\left(\ddd{F}{A_i}\cdot\left(\sum_{k=1}^3 c^k_{i,j}[A_k,\ddd{G}{A_j}]\right)+d_{i,j}[\sigma,\ddd{G}{A_j}]\right).
\]
Since $\sigma$ is a constant and the vector fields $Z_i$ admit the
matrix representation $Z_i(A_j)=\delta_{ij}Z(A)$ we see that the Lie
derivatives $Z_i(\{F,G\}_Q)$ is a multiple sum of terms like those of
Eq.~\rref{eq:zff0}, and so vanish whenever $Z(F)=Z(G)=0$. 
This proves that on the symplectic leaves of $P$ the \bih\
pencil $Q-\la P$ induces an $\omega N$ structure whose DN coordinates separate
the Hamilton--Jacobi equations of the Gaudin models.\\
{\bf Acknowledgments:} This work is an outgrowth of a long standing
research project of one of us (GF) with F. Magri and M. Pedroni.

\end{document}